\input{aipcheck}

\documentclass[final]{aipproc}

\layoutstyle{8x11single}

\begin{document}

\title{Charmonium molecules?}


\author{A. Valcarce}{
  address={Departamento de F\'\i sica Fundamental, Universidad de Salamanca, Salamanca, Spain.}
}

\author{J. Vijande}{
  address={Departamento de F\'\i sica At\'omica, Molecular y Nuclear, Universidad de Valencia (UV)
and IFIC (UV-CSIC), Valencia, Spain.}
}

\author{T. Fern\'andez-Caram\'es}{
  address={Departamento de F\'\i sica Fundamental, Universidad de Salamanca, Salamanca, Spain.}
}

\begin{abstract}
In this talk we present some recent 
studies of multiquark components in the charmonium sector.
We study the possible existence of compact four quark-states
and meson-meson molecules in the charmonium spectroscopy.
\end{abstract}

\maketitle

Since 2003 several states have been discovered in the charmonium mass region.
While in the conventional description
of the charmonium spectrum in terms of quark-antiquark pairs
some states are still missing, the number of experimental states
reported up to now is larger than empty spaces in the
$c\bar c$ spectrum. This, together with other difficulties
to explain observed states as simple quark-antiquark
pairs triggered discussions on a possible exotic interpretation:
four-quark states either as compact tetraquarks or slightly
bound meson-meson molecules.

Some states recently found in the hidden-charm sector~\cite{Ros07} may fit in the 
simple quark-model description as $(c\bar c)$ pairs 
(i.e., $X(3940)$, $Y(3940)$, and $Z(3940)$ 
as radially excited $\chi_{c0}$, $\chi_{c1}$, and $\chi_{c2}$), but
others appear to be more elusive, in particular $X(3872)$, $Z(4430)^+$, and $Y(4140)$. 
The debate on the nature of these states, including the possibility of finding 
$c\bar c n\bar n$ four-quark states, is open with special emphasis on the 
$X(3872)$. Since it was first reported by Belle in 2003~\cite{Bel03}, 
it has gradually become the flagship of the new armada of states 
whose properties make their identification as
traditional $(c\bar c)$ states unlikely.
An average mass of $3871.2\pm0.5\;$MeV and a narrow width of less than $2.3\;$MeV
have been reported for the $X(3872)$.
Note the vicinity of this state to the $D^0\overline{D}{}^{*0}$ threshold,
$M(D^0\,\overline{D}{}^{*0})=3871.2\pm1.2\;$MeV.
With respect to the $X(3872)$ quantum numbers, although some caution is still 
required until better statistic
is obtained~\cite{Set06}, an isoscalar $J^{PC}=1^{++}$ state seems to
be the best candidate to describe the properties of the $X(3872)$.  

In an attempt to disentangle the role played by multiquark configurations 
in the charmonium spectroscopy we have obtained an exact solution of the 
four-body problem based on an infinite expansion of the four-quark 
wave function in terms of hyperspherical harmonics (HH)~\cite{Vij07}.
The four-body  Schr\"odinger equation
has been solved accurately using two standard quark models
containing a linear confinement supplemented by a Fermi--Breit one-gluon exchange
interaction (BCN), and also boson exchanges
between the light quarks (CQC). The model parameters were tuned
in the meson and baryon spectra. 
The results are given in Table~\ref{tab:a}, indicating the quantum numbers of the
state studied, $J^{PC}$, the maximum value of the grand angular momentum used in the HH 
expansion, $K_{\rm max}$, and the
energy difference between the mass of the 
four-quark state, $E_{4q}$, and that of the lowest two-meson
threshold calculated with the same potential model, $\Delta_E$.

There are several issues that deserve our attention. First of all, the 
two-meson thresholds have been determined assuming quantum number conservation 
within exactly the same scheme used in the four--quark calculation. 
Dealing with strongly interacting particles, the two-meson states should have well defined total angular 
momentum, parity, and a properly symmetrized wave function if two identical mesons
are considered (coupled scheme). 
When noncentral forces are not taken into account, orbital angular 
momentum and total spin are also good quantum numbers (uncoupled scheme). 
Although noncentral forces were not used, the coupled scheme is the relevant one
for comparing with experimental data.

\begin{table}[b]
\begin{tabular}{ccccc} 
\hline
&\tablehead{2}{c}{b}{CQC} &\tablehead{2}{c}{b}{BCN} \\
\hline
\tablehead{1}{c}{b}{$J^{PC}(K_{max})$} &
\tablehead{1}{c}{b}{$E_{4q}$} &
\tablehead{1}{c}{b}{$\Delta_{E}$}&
\tablehead{1}{c}{b}{$E_{4q}$} & 
\tablehead{1}{c}{b}{$\Delta_{E}$} \\
\hline
$0^{++}$ (24) & 3779 &  +34 &  3249 &  +75  \\
$0^{+-}$ (22) & 4224 &  +64 &  3778 & +140  \\
$1^{++}$ (20) & 3786 &  +41 &  3808 & +153  \\
$1^{+-}$ (22) & 3728 &  +45 &  3319 &  +86  \\
$2^{++}$ (26) & 3774 &  +29 &  3897 &  +23  \\
$2^{+-}$ (28) & 4214 &  +54 &  4328 &  +32  \\
$1^{-+}$ (19) & 3829 &  +84 &  3331 & +157  \\
$1^{--}$ (19) & 3969 &  +97 &  3732 &  +94  \\
$0^{-+}$ (17) & 3839 &  +94 &  3760 & +105  \\
$0^{--}$ (17) & 3791 & +108 &  3405 & +172  \\
$0^{--}$ (17) & 3791 & +108 &  3405 & +172  \\
$2^{-+}$ (21) & 3820 &  +75 &  3929 &  +55  \\
$2^{--}$ (21) & 4054 &  +52 &  4092 &  +52 \\
\hline
\end{tabular}
\caption{$c\bar c n\bar n$ results.}
\label{tab:a}
\end{table}

\begin{figure}[t]
\vspace*{3cm}
        \includegraphics[height=.25\textheight]{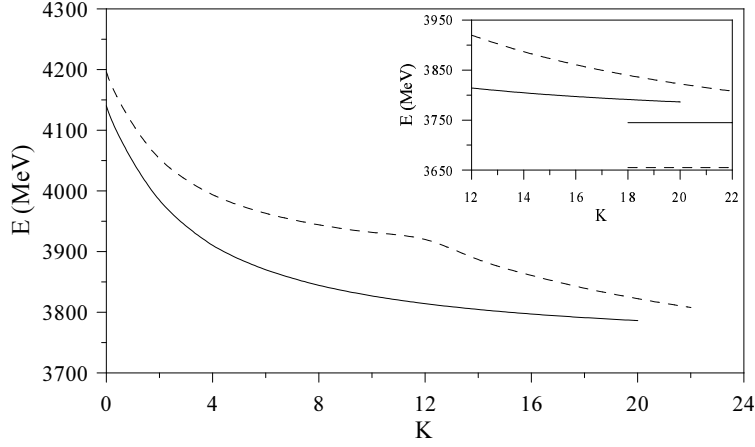}
\caption{Energy of the $1^{++}$ state using the CQC (solid line)
and BCN models (dashed line) as a function of $K$. The insert
in the upper-right corner magnifies the large values of $K$
to show the convergence to the corresponding threshold showed
by a straight line.}
\label{f1}
\end{figure}

Secondly, it is particularly interesting to check
the convergence of the method. We have plotted
in Fig.~\ref{f1} the energy of the $J^{PC}=1^{++}$ state
as a function of $K$. It can be observed how
the BCN $1^{++}$ state does not converge to the lowest
threshold for small values of $K$, being affected by the
presence of an intermediate $J/\psi\,\omega\vert_S$
threshold with an energy of 3874 MeV. Once sufficiently
large values of $K$ are considered the system follows
the usual convergence to the lowest threshold
(see insert in Fig.~\ref{f1}). Thus, the method may find difficulties
for states that are close below a two-meson threshold.

Finally, besides trying to unravel the possible existence of
bound $c\bar c n\bar n$ states
one should aspire to understand whether it is possible to differentiate between compact
and molecular states. A molecular state may be understood as a four-quark state
containing a single physical two-meson component, i.e., a unique singlet-singlet
component in the color wave function with well-defined spin and isospin quantum numbers.
One could expect these states not being deeply bound and therefore having a size
of the order of the two-meson system, i.e., 
$\Delta_R = R_{4q}/(r^1_ {2q}+r^2_{2q})\sim 1$. Opposite to that,
a compact state may be characterized by its involved structure on the
color space, its wave function containing different singlet-singlet
components with non negligible probabilities. One would expect
such states would be smaller than typical two-meson systems, i.e.,
$\Delta_R < 1$. Let us notice that while $\Delta_R>1$ but finite would 
correspond to a meson-meson molecule 
$\Delta_R\stackrel{K\to\infty}{\longrightarrow}\infty$
would represent a two-meson threshold.

Considering these remarks and the results shown in Table~\ref{tab:a} one can conclude
that there are no deeply bound states (compact) in the $c\bar c n\bar n$ system.
However, as explained above, being the HH method exact, it is not completely 
adequate to study states that are close to,
but below, the charmed meson production threshold.
Such states are called {\it molecular}, in the sense
that they can be exactly expanded in terms of a single singlet-singlet
color vector. 
Close to a threshold, methods based on a series expansion may fail to converge
since arbitrary large number of terms are required to determine the wave function.

As mentioned above, the new experimental findings do not fit, in general, the simple predictions of the quark-antiquark
schemes and, moreover, they overpopulate the expected number of states in (simple) two-body theories.
This situation is not uncommon in particle physics. For example,
in the light scalar-isoscalar meson sector hadronic molecules seem to
be needed to explain the experimental data~\cite{Jaf77,Wei90,Hoo08}. Also,
the study of the $NN$ system above the pion production threshold 
required new degrees of freedom to be incorporated in the theory, either
as pions or as excited states of the nucleon, i.e., 
the $\Delta$~\cite{Els88,Pop87}. 
This discussion suggests that charmonium spectroscopy could be rather
simple below the threshold production of charmed mesons but  
much more complex above it. In particular, the coupling to 
the closest $(c\bar c)(n\bar n)$ system, referred to 
as {\it unquenching the naive quark model}~\cite{Clo05},
could be an important spectroscopic ingredient. Besides,
hidden-charm four-quark states could 
explain the overpopulation of quark-antiquark theoretical states.

Trying to look for the possible existence of loosely bound molecular states close
to a charmed meson threshold, we have used a different technique~\cite{Fer09},
we have solved the Lippmann-Schwinger equation 
looking for attractive channels that may contain a meson-meson molecule.
In order to account for all basis states we allow 
for the coupling to charmonium-light two-meson systems.

When we consider the system of two mesons $M_1$ and $\overline{M}_2$ ($M_i=D, D^*$) 
in a relative $S-$state interacting through a potential $V$ that contains a
tensor force then, in general, there is a coupling to the 
$M_1\overline{M}_2$ $D-$wave and the
Lippmann-Schwinger equation of the system is 
\begin{equation}
t_{ji}^{\ell s\ell^{\prime \prime }s^{\prime \prime }}(p,p^{\prime \prime };E)
=V_{ji}^{\ell s\ell^{\prime \prime }s^{\prime \prime }}(p,p^{\prime \prime
})+\sum_{\ell^{\prime }}\int_{0}^{\infty }{p^{\prime }}%
^{2}dp^{\prime }
\, V_{ji}^{\ell s \ell^{\prime }s^{\prime}}
(p,p^{\prime })
{\frac{1}{E-{p^{\prime }}^{2}/2{\bf \mu }+i\epsilon }}%
t_{ji}^{\ell^{\prime }s^{\prime }\ell^{\prime \prime }s^{\prime \prime
}}(p^{\prime },p^{\prime \prime };E),  \label{eq1}
\end{equation}
where $t$ is the two-body amplitude, $j$, $i$, and $E$ are the
angular momentum, isospin and energy of the system, and $\ell s$,
$\ell^{\prime }s^{\prime }$, $\ell^{\prime \prime }s^{\prime \prime }$
are the initial, intermediate, and final orbital angular momentum
and spin; $p$
and $\mu $ are the relative momentum and reduced mass of the
two-body system, respectively. 
In the case of a two $D$ meson system that can couple to a charmonium-light
two-meson state, for example when $D\overline{D}^*$ is coupled to
 $J/\Psi \omega$, the Lippmann-Schwinger equation for
$D\overline{D}^*$ scattering becomes
\begin{eqnarray}
t_{\alpha\beta;ji}^{\ell_\alpha s_\alpha \ell_\beta s_\beta}(p_\alpha,p_\beta;E) = 
V_{\alpha\beta;ji}^{\ell_\alpha s_\alpha \ell_\beta s_\beta}(p_\alpha,p_\beta)&+&
\sum_{\gamma}\sum_{\ell_\gamma} 
\int_0^\infty p_\gamma^2 dp_\gamma
V_{\alpha\gamma;ji}^{\ell_\alpha s_\alpha \ell_\gamma s_\gamma}
(p_\alpha,p_\gamma) \nonumber \\
&\times& \, G_\gamma(E;p_\gamma)
t_{\gamma\beta;ji}^{\ell_\gamma s_\gamma \ell_\beta s_\beta}
(p_\gamma,p_\beta;E),
\label{eq2}
\end{eqnarray}
with $\alpha, \beta, \gamma= D\overline{D}^*, J/\Psi \omega$.

We have consistently used the same interacting Hamiltonian to study
the two- and four-quark systems to guarantee that thresholds
and possible bound states are eigenstates of the same Hamiltonian. 
Such interaction contains a universal
one-gluon exchange, confinement, and a chiral potential between
light quarks~\cite{Vij05}. We have solved the coupled channel problem of the 
$D\overline D$, $D\overline{D}^*$, and $D^*\overline{D}^*$. In all cases
we have included the coupling to the relevant $(c\bar c)(n\bar n)$
channel (from now on denoted as $J/\Psi \omega$ channels).

As we study systems with well-defined 
$C-$parity and
since neither $D\overline{D}^*$ nor $\overline{D}D^*$ 
are eigenstates of $C-$parity, it
is necessary to construct the proper linear combinations. Taking into account that
$C(D) = \overline{D}$ and $C(D^*) = - \overline{D}^*$, it can be found that~\cite{Bra07}:
\begin{equation}
D_1 = \frac{1}{\sqrt{2}} \left( D\overline{D}^* + \overline{D}D^*\right)
\end{equation}
and
\begin{equation}
D_2 = \frac{1}{\sqrt{2}} \left( D\overline{D}^* - \overline{D}D^*\right)
\end{equation}
are the eigenstates corresponding to $C = -1$ and $C = +1$, respectively.

\begin{figure}[t]
  \includegraphics[height=.3\textheight]{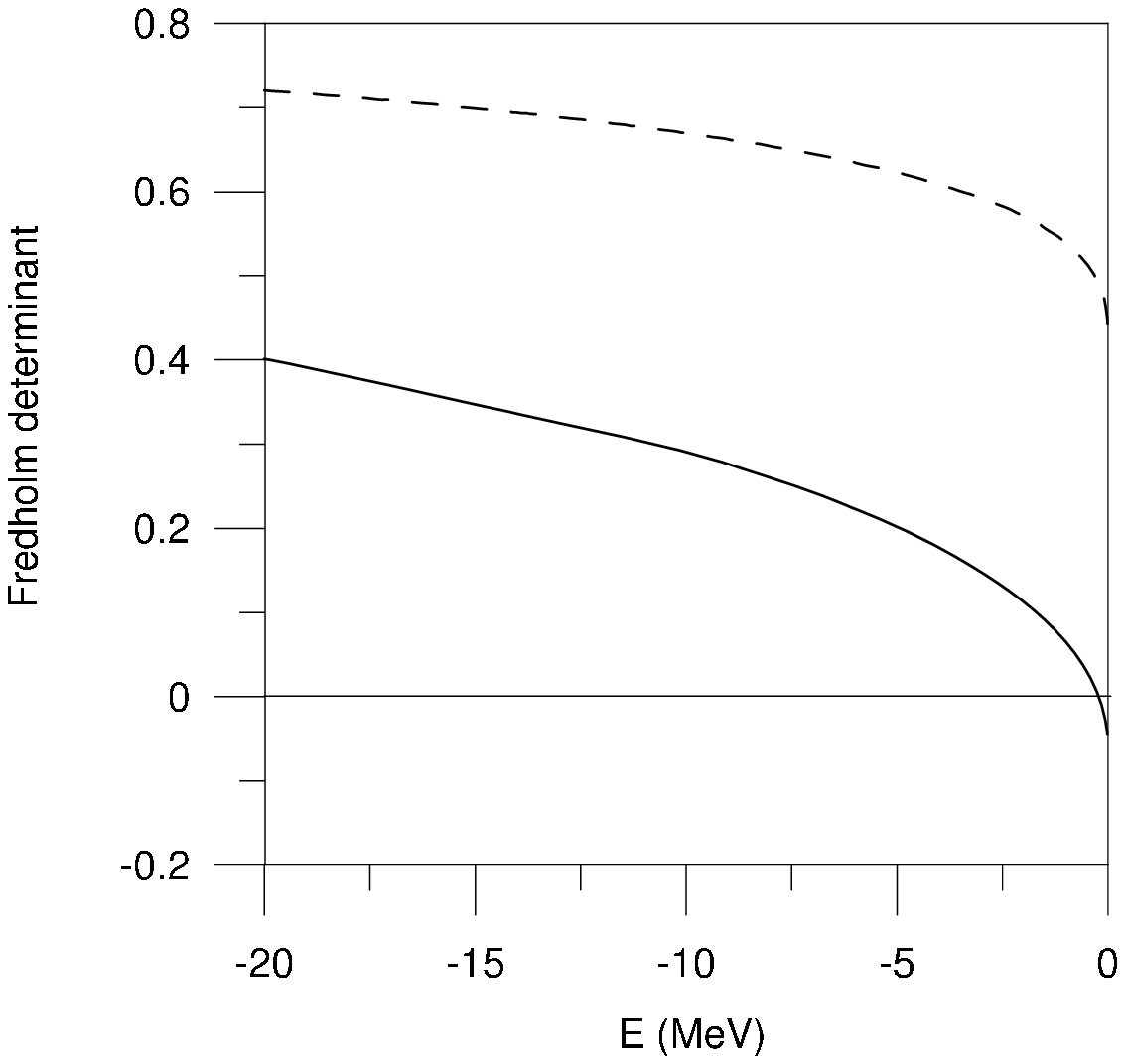}
\caption{Fredholm determinant for the $J^{PC}(I)=1^{++}(0)$ $D\overline{D}^*$
system. Solid (dashed) line: results with (without) coupling to the $J/\Psi \omega$
channel.}
\label{f2}
\end{figure}

Table~\ref{t1} and Fig.~\ref{f2} summarize our results.
We have specified the quantum numbers of the attractive channels. 
The rest, not shown on the table, are either repulsive or have zero probability to 
contain a bound state or a resonance.
Let us remark that, of all possible channels, only a few are attractive.
Of the systems made of a particle and its corresponding antiparticle,
the $J^{PC}(I)=0^{++}(0)$ channel is always attractive. In general,
the coupling to the $\eta_c \eta$ channel
reduces the attraction, but there is still enough attraction to expect a resonance 
close and above the threshold. This channel is much more attractive
for the $D^*\overline{D}^*$ system than for $D\overline{D}$, thus, in the latter 
one could expect a wider resonance. 
It is easy to explain the reason for such a close-to-bind situation 
with these quantum numbers. They
can be reached from a two-meson system without explicit orbital
angular momentum, while through a simple $c\bar c$ pair
it needs a unit of orbital angular momentum. 
Similar arguments were used to explain the proliferation
of light scalar-isoscalar mesons ~\cite{Jaf77,Wei90,Hoo08}.
The most attractive channel in the $D\overline{D}^*$ case
is the $J^{PC}(I)=1^{++}(0)$ and can be explained as before, except 
the unity of intrinsic spin due to the $D^*$ meson.
A simple calculation of the $D\overline{D}^*$ system (Eq.~(\ref{eq1})) indicates that
the $J^{PC}(I)=1^{++}(0)$ and $1^{+-}(1)$ are degenerate. It is
the coupling to the $J/\Psi \omega$ (Eq.~(\ref{eq2})) that breaks the degeneracy to
make the $1^{++}(0)$ more attractive. 
The isospin 1
channel becomes repulsive due to the coupling to the lightest channel that 
includes a pion. Then, the existence of
meson-meson molecules in the isospin one $D\overline{D}^*$ channels can be discarded.
Using the coupling to the $J/\Psi \omega$, not present
in the calculations at the hadronic level of~\cite{Liu08,Tho08},
we obtain a
binding energy for the $J^{PC}(I)=1^{++}(0)$ in the range $0-1$ MeV,
in good agreement with the experimental measurements of $X(3872)$ (see Fig.~\ref{f2}).
This result supports
the analysis of the Belle data on $B \to K + J/\Psi \pi^+ \pi^-$ and
$B \to K + D^0 \overline{D}^0 \pi^0$ that favors the $X(3872)$ being
a bound state whose mass is below the $D^0\overline{D}^0$ 
threshold~\cite{Bra07}. The existence of a bound state
in the $1^{++}(0)$ $D\overline{D}^*$ channel would not show up
in the $D\overline{D}$ system because of quantum
number conservation.

Finally, we have found that the $J^{PC}(I)=2^{++}(0,1)$ $D^*\overline{D}^*$
are also attractive due to the coupling to the $J/\Psi \omega$ and 
$J/\Psi \rho$ channels, respectively. This would give rise to new
states around 4 GeV/c$^2$ and one 
experimental candidate could be the $Y(4008)$.
In this case, such a resonance would also appear
in the $D\overline{D}$ system for large relative
orbital angular momentum, $L=2$. A
similar behavior can be observed in resonances predicted
for the $\Delta \Delta$ system~\cite{Val01}.
\begin{table}[b]
\begin{tabular}{cc}
\hline
\tablehead{1}{c}{b}{System} & \tablehead{1}{c}{b}{$J^{PC}(I)$} \\
\hline
$D\overline{D}$ & $0^{++}(0)$ \\ 
\hline
$D\overline{D}^*$ & $1^{++}(0)$ \\ 
\hline
$D^*\overline{D}^*$ & $0^{++}(0)$ \\ 
$D^*\overline{D}^*$ & $2^{++}(0)$ \\ 
$D^*\overline{D}^*$ & $2^{++}(1)$ \\ 
\hline
\end{tabular}
\caption{Attractive channels for the two $D-$mesons system.}
\label{t1}
\end{table}
 
In all cases, being loosely bound states whose masses are
close to the sum of their constituent meson masses,
their decay and production properties must be quite 
different from conventional $q\bar q$ mesons.
Our calculation does not exclude a possible mixture of
standard charmonium states in the channels where
we have found attractive molecular systems. 
This admixture could explain some properties of the
$X(3872)$~\cite{Ger06,Big09}. 
We would like to emphasize the similarity of our results
to those of Ref.~\cite{Tor91} in spite 
of our different approach. Our treatment is general, dealing
simultaneously with the two- and four-body problems and using
an interaction containing gluon and quark exchanges instead of
the simple two-body one-pion exchange potential of Ref.~\cite{Tor91}.
Nevertheless, we also concluded that
the lighter meson-meson molecules are in
the vector-vector and pseudoscalar-vector two-meson channels. 
Finally, let us remark that our approach could also be applied to the
the $c\bar cs\bar s$ sector.

To summarize, our predictions show that no deeply bound states can be expected for 
the $c\bar c n\bar n$ system.
Only a few channels can be expected to 
present observable resonances or slightly bound states. Among them, we
have found that the $D\overline{D}^*$ system must show a bound state
slightly below the threshold for charmed mesons production
with quantum numbers $J^{PC}(I)=1^{++}(0)$,
that could correspond to the widely discussed $X(3872)$. Of the systems 
made of a particle and its corresponding antiparticle,
$D\overline{D}$ and $D^*\overline{D}^*$, the $J^{PC}(I)=0^{++}(0)$
is attractive. It would be the only candidate to accommodate
a wide resonance for the $D\overline{D}$ system. 
For the $D^*\overline{D}^*$ the attraction
is stronger and structures may be observed close and above the charmed
meson production threshold.
Also, we have shown that the $J^{PC}(I)=2^{++}(0,1)$ $D^*\overline{D}^*$
channels are attractive due to the coupling to the $J/\Psi \omega$ and 
$J/\Psi \rho$ channels. 
Due to heavy quark symmetry, replacing the charm
quarks by bottom quarks decreases the kinetic energy without significantly
changing the potential energy. In consequence, four-quark bottomonium mesons
must also exist and have larger binding energies.

Particular analysis of different states based on different techniques have arrived 
to similar conclusions: Ref.~\cite{Gut09}, based on effective lagrangians, concludes
that the $Y(3940)$ could be a $D^* \overline D^*$ $J^{PC}(I)=0^{++}(0)$ or $2^{++}(0)$ meson-meson molecule and
the $Y(4140)$ could be a $D_s^* \overline D_s^*$ $J^{PC}(I)=0^{++}(0)$ or $2^{++}(0)$ meson-meson molecule;
Ref.~\cite{Mol09}, based on dynamically generated resonances, concludes that the $Y(3940)$,
$Z(3940)$ and $X(4160)$ could be $D^* \overline D^*$ and $D_s^* \bar D_s^*$ $0^{++}(0)$ and
$2^{++}(0)$ states; Ref.~\cite{Alb09}, based on QCD sum rules, concludes
that the $Y(4140)$ could be a $D_s^* \overline D_s^*$ $J^{PC}(I)=0^{++}(0)$ or $2^{++}(0)$ meson-meson molecule;
Ref.~\cite{Din09}, based on a one-boson exchange model, concludes
that the $Y(4140)$ could be a $D_s^* \overline D_s^*$ $J^{PC}(I)=0^{++}(0)$ meson-meson molecule.

\begin{theacknowledgments}
This work has been partially funded by the Spanish Ministerio de
Educaci\'on y Ciencia and EU FEDER under Contract No. FPA2007-65748,
by Junta de Castilla y Le\'{o}n under Contract No. GR12, 
and by the Spanish Consolider-Ingenio 2010 Program CPAN (CSD2007-00042),
\end{theacknowledgments}

\end{document}